\documentstyle[]{article}
\def\cn{\hbox{cn}}
\def\sn{\hbox{sn}}
\def\dn{\hbox{dn}}

\def\bea{\begin{eqnarray}}
\def\eea{\end{eqnarray}}
\def\nn{\nonumber}
\def\beq{\begin{equation}}
\def\eeq{\end{equation}}
\def\ba{\beq\new\begin{array}{c}}
\def\ea{\end{array}\eeq}
\def\be{\ba}
\def\ee{\ea}

\makeatletter
\newdimen\normalarrayskip % skip between lines
\newdimen\minarrayskip % minimal skip between lines
\normalarrayskip\baselineskip
\minarrayskip\jot
\newif\ifold \oldtrue \def\new{\oldfalse}
\def\arraymode{\ifold\relax\else\displaystyle\fi} % mode of array entries
\def\eqnumphantom{\phantom{(\theequation)}} % right phantom in eqnarray
\def\@arrayskip{\ifold\baselineskip\z@\lineskip\z@
\else
\baselineskip\minarrayskip\lineskip2\minarrayskip\fi}
\def\@arrayclassz{\ifcase \@lastchclass \@acolampacol \or
\@ampacol \or \or \or \@addamp \or
\@acolampacol \or \@firstampfalse \@acol \fi
\edef\@preamble{\@preamble
\ifcase \@chnum
\hfil$\relax\arraymode\@sharp$\hfil
\or $\relax\arraymode\@sharp$\hfil
\or \hfil$\relax\arraymode\@sharp$\fi}}
\def\@array[#1]#2{\setbox\@arstrutbox=\hbox{\vrule
height\arraystretch \ht\strutbox
depth\arraystretch \dp\strutbox
width\z@}\@mkpream{#2}\edef\@preamble{\halign
\noexpand\@halignto
\bgroup \tabskip\z@ \@arstrut \@preamble \tabskip\z@ \cr}%
\let\@startpbox\@@startpbox \let\@endpbox\@@endpbox
\if #1t\vtop \else \if#1b\vbox \else \vcenter \fi\fi
\bgroup \let\par\relax
\let\@sharp##\let\protect\relax
\@arrayskip\@preamble}
\def\eqnarray{\stepcounter{equation}%
\let\@currentlabel=\theequation
\global\@eqnswtrue
\global\@eqcnt\z@
\tabskip\@centering
\let\\=\@eqncr
$$%
\halign to \displaywidth\bgroup
\eqnumphantom\@eqnsel\hskip\@centering
$\displaystyle \tabskip\z@ {##}$%
\global\@eqcnt\@ne \hskip 2\arraycolsep
$\displaystyle\arraymode{##}$\hfil
\global\@eqcnt\tw@ \hskip 2\arraycolsep
$\displaystyle\tabskip\z@{##}$\hfil
\tabskip\@centering
&{##}\tabskip\z@\cr}
\begingroup\ifx\undefined\newsymbol \else\def\input#1 {\endgroup}\fi
\newfont{\hr}{msbm10}
\newfont{\ams}{msam10}
\font\numbers=cmss12
\font\upright=cmu10 scaled\magstep1
\def\stroke{\vrule height8pt width0.4pt depth-0.1pt}
\def\topfleck{\vrule height8pt width0.5pt depth-5.9pt}
\def\botfleck{\vrule height2pt width0.5pt depth0.1pt}
\def\Zmath{\vcenter{\hbox{\numbers\rlap{\rlap{Z}\kern 0.8pt\topfleck}\kern
2.2pt \rlap Z\kern 6pt\botfleck\kern 1pt}}}
\def\Qmath{\vcenter{\hbox{\upright\rlap{\rlap{Q}\kern
3.8pt\stroke}\phantom{Q}}}}
\def\Nmath{\vcenter{\hbox{\upright\rlap{I}\kern 1.7pt N}}}
\def\Cmath{\vcenter{\hbox{\upright\rlap{\rlap{C}\kern
3.8pt\stroke}\phantom{C}}}}
\def\Rmath{\vcenter{\hbox{\upright\rlap{I}\kern 1.7pt R}}}
\def\Z{\ifmmode\Zmath\else$\Zmath$\fi}
\def\Q{\ifmmode\Qmath\else$\Qmath$\fi}
\def\N{\ifmmode\Nmath\else$\Nmath$\fi}
\def\C{\ifmmode\Cmath\else$\Cmath$\fi}
\def\R{\ifmmode\Rmath\else$\Rmath$\fi}
\newcounter{app}

\def\app{\setcounter{equation}{0}
\def\theequation{\Alph{app}.\arabic{equation}}\par
\addvspace{4ex}
\@afterindentfalse
\secdef\@app\@dapp}

\newcommand\@app{\@startsection {app}{1}{0ex}%
{-3.5ex \@plus -1ex \@minus -.2ex}%
{2.3ex \@plus.2ex}%
{\normalfont\Large\bf}}
\def\@dapp#1{%
{\parindent \z@ \raggedright \bf #1}\par\nobreak}
\def\l@app#1#2{\ifnum \c@tocdepth >\z@
\addpenalty\@secpenalty
\addvspace{1.0em \@plus\p@}%
\setlength\@tempdima{8em}%
\begingroup
\parindent \z@ \rightskip \@pnumwidth
\parfillskip -\@pnumwidth
\leavevmode \bfseries
\advance\leftskip\@tempdima
\hskip -\leftskip
#1\nobreak\hfil \nobreak\hb@xt@\@pnumwidth{\hss #2}\par
\endgroup\fi}
\newcounter{sapp}[app]

\def\sapp{\def\theequation{\Alph{app}.\arabic{equation}}
\par
\@afterindentfalse
\secdef\@sapp\@dsapp}
\newcommand{\@sapp}{\@startsection{sapp}{2}{\z@}%
{-3.25ex\@plus -1ex \@minus -.2ex}%
{1.5ex \@plus .2ex}%
{\normalfont\large\bfseries}}

\def\@dsapp#1{%
{\parindent \z@ \raggedright \bf #1}\par\nobreak}
\newcommand{\l@sapp}{\@dottedtocline{2}{1.5em}{2.3em}}

\def\2{{1\over 2}}
\def\N2{${\cal N}=2$}

\def\be{ \begin{eqnarray} }
\def\ee{ \end{eqnarray} }

\def\g{\sqrt{-2g^2}}
\def\bea{\begin{eqnarray}}
\def\eea{\end{eqnarray}}
\def\nn{\nonumber}

\def\beq{\begin{equation}}
\def\eeq{\end{equation}}
\def\ba{\beq\new\begin{array}{c}}
\def\ea{\end{array}\eeq}
\def\be{\ba}
\def\ee{\ea}

\def\R{Ruijsenaars\ }

\begin{document}
\begin{flushright}
\hfill ITEP/TH-24/99\\
\hfill FIAN/TD-06/99\\
hepth/9906240
\end{flushright}
\vspace{0.5cm}
\begin{center}
{\Large\bf On Double-Elliptic Integrable Systems}\\
{\Large\bf 1. A Duality Argument for the case of SU(2)}
\vspace{0.5cm}

\setcounter{footnote}{1}
\def\thefootnote{\fnsymbol{footnote}}
{\Large H.W.Braden\footnote{Department of Mathematics and Statistics,
University of Edinburgh, Edinburgh EH9 3JZ Scotland;
e-mail address: hwb@ed.ac.uk},
A.Marshakov\footnote{Theory
Department, Lebedev Physics Institute, Moscow
~117924, Russia; e-mail address: mars@lpi.ac.ru}\footnote{ITEP,
Moscow 117259, Russia;
e-mail address:
andrei@heron.itep.ru},
A.Mironov\footnote{Theory
Department, Lebedev Physics Institute, Moscow
~117924, Russia; e-mail address: mironov@lpi.ac.ru}\footnote{ITEP,
Moscow 117259, Russia; e-mail address:
mironov@itep.ru}, A.Morozov\footnote{ITEP, Moscow
117259, Russia; e-mail address: morozov@vx.itep.ru}
}\\
\end{center}
\bigskip
\begin{quotation}
We construct a two parameter family of 2-particle Hamiltonians
closed under the duality operation of
interchanging the (relative) momentum and coordinate.
Both coordinate and momentum dependence are elliptic,
and the modulus of the momentum torus is a non-trivial
function of the coordinate.
This model contains as limiting cases the standard
Ruijsenaars-\-Calogero and Toda family of Hamiltonians,
which are at most elliptic in the coordinates, but not in the momenta.
\end{quotation}

%\bigskip

\section{Introduction}
\setcounter{footnote}{0}
%{\bf 1.}
The theory of (classical) integrable systems has been the subject of renewed
interest following the realisation that integrability is a crucial
and characteristic property of non-perturbative effective actions
(see, for example, \cite{effac}).
From this point of view the low-energy effective actions of
Yang-Mills theories \cite{SW,SWgen}
(they are non-trivial, for example, in the
presence of $N=2$ supersymmetry) belong to universality classes represented
by the simplest finite-dimensional integrable models \cite{GKMMM,DW,rev}.
At the same time these
Yang-Mills theories may be associated with D-branes \cite{branes},
which can be embedded in various target spaces.
Unfortunately the set of known integrable models (the
Calogero-\-Ruijsenaars and Toda family\footnote{
The (second) Hamiltonian of the Calogero system \cite{Cal,Krisol} is
$$H_2^{Cal} = \sum_i \frac{p_i^2}{2} +
\sum_{i<j} V(q_{ij})$$
($q_{ij} = q_i - q_j$ and in the centre-of-mass $H_1^{Cal} =
\sum_i p_i = 0$),
with $V(q)$ a rational, trigonometric or elliptic
function of the coordinates with second order pole.
The conserved Hamiltonians of the Calogero
family may be written so as to exhibit a rational (polynomial)
dependence on the momenta.

The (first) Ruijsenaars Hamiltonian \cite{Ru} is
$$H_1^{Ru} = \sum_i \cosh p_i \prod_{k\neq i}F(q_{ik})$$
where $F(q)$ is a rational, trigonometric or an elliptic function.
The momentum dependence of the conserved Hamiltonians
is now trigonometric, with the Calogero family arising as a
limit of the Ruijsenaars one.

The Toda-chain family \cite{Toda} (which warrants a special mention
because it is associated with pure gauge $N=2$ SUSY Yang-Mills models
in 4 dimensions \cite{GKMMM,MW,NT})
is a special double-scaling limit of the elliptic
Calogero model \cite{Ino}.

Each of these models, while being various limits of the elliptic
Ruijsenaars system, may
further be embedded into the
double-elliptic system introduced in the present paper.
})
does not include all of the universality classes arising from the
various brane constructions.
The main gap is a putative ``double-elliptic'' integrable
system, where both coordinates and momenta take values
in elliptic curves (complex tori), which should play
a role in the description of toric, K3 and Calabi-Yau
target spaces. These will be associated with (compactified) six-dimensional
Yang-Mills theories.

It is the task of this paper to suggest what such a
double-elliptic system can look like.
We will discuss the most straightforward construction based on a
duality argument for the case of Yang-Mills gauge group $SU(2)$
(2-particle integrable system).
By itself our argument is not conclusive, for in this situation there
is only one non-trivial Hamiltonian and the multiparticle generalisation
is needed.
Nonetheless, it provides an important insight.

Our paper first reviews duality in the context of integrable systems.
Using this, section three constructs a Hamiltonian dual to the
(elliptic) Calogero model that is elliptic in the momentum.
The resulting ``rational-elliptic" model is investigated with
comparison made both to a direct solution of the elliptic Calogero model
and that resulting from use of the ``projection"
method of Olshanetsky and Perelomov.
An interesting feature of our model is the appearance of ``dressed"
elliptic curves. Section four simply states the result for the
dual of the (elliptic) Ruijsenaars model, our ``trigonometric-elliptic"
model, while section five details the construction of the wholly
new ``elliptic-elliptic" model. We show here how the Ruijsenaars-Calogero
and Toda families arise as limits of our model. Some comments are also
made on the ansatz involved in the construction of our model. We end
with a brief conclusion.

\section{Duality of integrable systems}

%\subsection{Approaches to integrability}

Integrable systems may be introduced and solved in a variety
of ways. Some of these include
\begin{enumerate}

\item The projection method \cite{OP}, where solvable and often
trivial dynamics on a given space look non-trivial after dimensional
reduction or projection to a lower dimensional space.

\item A full list of Hamiltonians in involution may be given.

\item The system is exhibited in the Lax form (possibly with spectral
parameter) and R-matrices given \cite{HWB},
showing the Poisson commutativity of the
powers of the traces. In field theory applications the Lax representation may
be deduced either from (possibly quantum) group theory, or from the dynamics
of scalars -- describing the shape of branes embedding \cite{branes} -- in
higher-dimensional SUSY Yang-Mills models. The latter is essentially the
DKN-Hitchin approach \cite{DKN,Hit}.

\item Generalized WDVV equations \cite{WDVV,WDVV3,Ma,WDVVrev}.
In the simplest cases they are related to Hodge theory \cite{D}
and, in more interesting situations,
to non-trivial algebras of forms \cite{WDVV3}.

\item Coordinate-momentum duality.

\end{enumerate}

In what follows we shall introduce and exploit the last approach,
which is quite a constructive procedure in the case of $SU(2)$.
Connection will also be made with the projection method and, simply because
of dimensionality, the Hamiltonian we construct together with the
centre-of-mass
fully describes the system. In this section we will
first explain the general idea behind ``duality" and
then apply it in the two-particle
context. We conclude the section by relating this duality
to an underlying DKN-Hitchin-Seiberg-Witten structure.

\subsection{Duality: the general idea}

The idea of duality here expresses a relationship between two completely
integrable systems $S_1, S_2$ on a fixed symplectic manifold with given
symplectic structure $(M,\omega)$ and goes back to
\cite{ruijsenaars,duality}.
We say the Hamiltonian systems are {\it dual} when the conserved quantities of
$S_1$ and $S_2$ together form a coordinate system for $M$. Consider for
example free particles, $H^{(1)}_k=\sum_i p^k_i/k$. For this system the free
particles momenta are identical to the conserved quantities or action
variables. Now consider the Hamiltonian $H^{(2)}_k=\sum_i q^k_i/k$ with
conserved quantities $q_i$. Together $\{p_i,q_i\}$ form a coordinate system
for phase space, and so the two sets of Hamiltonians are dual. Duality then
in this simplest example is a transformation which interchanges momenta and
coordinates. For more complicated interacting integrable systems finding dual
Hamiltonians is a nontrivial exercise. Note that this whole construction
manifestly depends on the particular choice of conserved quantities. A
clever choice may result in the dual system arising by simply interchanging
the momentum and coordinate dependence, as in the free system.

Some years ago Ruijsenaars \cite{ruijsenaars} observed such
dualities between various members of the Calogero-Moser and Ruijsenaars
families: the rational Calogero and trigonometric Ruijsenaars models were
dual to themselves while trigonometric Calogero model was dual with the
rational Ruijsenaars system (see \cite{duality} for more examples).
These dualities were shown by starting with a
Lax pair $L=L(p,q)$ and an auxillary diagonal matrix $A=A(q)$. When $L$ was
diagonalized the matrix $A$ became the Lax matrix for the dual Hamiltonian,
while $L$ was a function of the coordinates of the dual system. Dual systems for
a model possessing a Lax representation are then related to the eigenvalue
motion of the Lax matrix.

Our approach to finding a dual system \cite{duality} is to make a canonical
transformation which
substitutes the original set of Poisson-commuting coordinates $q_i$,
$\{q_i,q_j\} = 0$, by another obvious
set of the Poisson-commuting
variables: the Hamiltonians $h_i(\vec p,\vec q)$ or, better, the
action variables $a_i(\vec h) = a_i(\vec p,\vec q)$.
It will be clear below that in practice really interesting
transformations are a little more sophisticated: $h_i$ are identified with
certain functions of the new coordinates (these functions determine the
Ruijsenaars matrix $A(q)$), which -- in the most interesting cases --
are just the same Hamiltonians with the interactions switched-off. Such free
Hamiltonians are functions of momenta alone, the and dual coordinates substitute
these momenta, just as one had for the system of free particles.

The most interesting question for our purposes is: {\it what are the duals of
the elliptic Calogero and Ruijsenaars systems ?}
Since the elliptic Calogero (Ruijsenaars) is rational
(trigonometric) in momenta and elliptic in the coordinates, the dual
will be elliptic in momenta
and rational (trigonometric) in coordinates.
Having found such a model the final elliptization of the coordinate
dependence is straightforward, providing us with the wanted
double-elliptic systems.

\subsection{The 2-particle ($SU(2)$) case}

The calculations are especially simple in the case of $SU(2)$
which, in the center-of-mass frame, has only one coordinate and
one momentum.
In this case the duality transformation can be described explicitly since
the equations of motion can be integrated in a straightforward way.
Technically, given two Hamiltonian systems, one with the momentum $p$,
coordinate $q$ and
Hamiltonian $h(p,q)$ and another with the momentum $P$,
coordinate $Q$ and
Hamiltonian $H(P,Q)$ we may describe duality by the relation

\be
h(p,q) = f(Q), \nn \\
H(P,Q) = F(q).
\ee
Here the functions $f(Q)$ and $F(q)$ are such that

\be
dP\wedge dQ = -dp\wedge dq,
\ee
which expresses the fact we have a canonical transformation. This
relation entails that

\be
F'(q)\frac{\partial h(p,q)}{\partial p} =
f'(Q)\frac{\partial H(P,Q)}{\partial P}.
\label{pbconseq}
\ee

At this stage the functions $f(Q)$ and $F(q)$ are arbitrary.
However, when the Hamiltonians depend on a coupling constant $g^
2$
and are such that their ``free'' part can be separated and depends
only on the momenta,\footnote{
Note, that this kind of duality relates the weak coupling
regime for $h(p,q)$ to the weak coupling regime for
$H(P,Q)$. For example, in the rational Calogero case
$$h(p,q) = \frac{p^2}{2} + \frac{g^2}{q^2} = \frac{Q^2}{2},$$
$$H(P,Q) = \frac{P^2}{2} + \frac{g^2}{Q^2} = \frac{q^2}{2}.$$
We recall that in the brane picture the coupling constant
$g$ is related to the mass of adjoint hypermultiplet and
thus remains unchanged under $T$-duality transformations.
}
the free Hamiltonians provide a natural choice for these functions:
$F(q) = h_0(q)$ and $f(Q) = H_0(Q)$ where
\be
h(p,q)\big|_{g^2 = 0} = h_0(p), \\
H(P,Q)\big|_{g^2 = 0} = H_0(P).
\ee
With such choice the duality equations become
\be
h_0(Q) = h(p,q) \label{duality1},\\
H_0(q) = H(P,Q) \label{duality2},\\
\frac{\partial h(p,q)}{\partial p}H'_0(q) =
h'_0(Q)\frac{\partial H(P,Q)}{\partial P}.
\label{duality3}
\ee

Free rational,
trigonometric and elliptic Hamiltonians are
$h_0(p) = \frac{p^2}{2}$, $h_0(p) = \cosh p$ and $h_0(p) =\cn(p|k)$
respectively.

\subsection{The DKN-Hitchin-Seiberg-Witten structure \label{HSW}}

The main duality relation,
\be
H_0(q) = H(P,Q)
\label{duality}
\ee
can be considered as defining a family of spectral curves
$q(P)$, parameterised by a parameter (modulus) $Q$.
The symplectic structure $dP\wedge dQ = - dp\wedge dq$, used in the formulation
of duality, is then related to the generating \lq\lq Seiberg-Witten" 1-form
$dS = pdq$, \footnote{\label{SWstr}
In the general case of a $g$-parameter family of complex curves
(Riemann surfaces) of genus $g$, the Seiberg-Witten differential
$dS$ is characterised by the property
$\delta dS = \sum_{i=1}^g \delta u_i dv_i$,
where $dv_i(z)$ are the $g$ holomorphic 1-differentials
on the curves (on the fibers), while $\delta u_i$ are the
variations of $g$ moduli (along the base).
The associated integrable system has $u_i$ as coordinates
and $\pi_i$
-- some $g$ points on the curve -- as the momenta.
The symplectic structure is
$$\sum_{i=1}^g da_i\wedge dp_i = \sum_{i,k=1}^g
du_i\wedge dv_i(\pi_k)$$
The vector $p_i = \sum_{k=1}^g \int^{\pi_k} d\omega_i$
is a point of the Jacobian, and the Jacobi map identifies this with the
$g$-th power of the curve, $Jac\ \cong {\cal C}^{\otimes g}$.
Here $d\omega_i$ are {\it canonical} holomorphic differentials,
$dv_i = \sum_{j=1}^g d\omega_j \oint_{A_j} dv_i$.
The Seiberg-Witten integrals
$$a_i(u) = \oint_
{A_i}dS,$$
which satisfy
$$\frac{\partial a_i}{\partial u_j} = \oint_{A_i}dv_j,$$
define a {\it flat} structure on the moduli space.
The generalized WDVV equations are written in terms of these coordinates
\cite{WDVV,WDVV3}. }.

From (\ref{duality}) it follows that
\be
\left.\frac{\partial q}{\partial P}\right|_Q = \frac{1}{H_0'(q)}
\frac{\partial H(P,Q)}{\partial P}
\ee
which together with (\ref{duality3}) implies:
\be
\left.\frac{\partial q}{\partial P}\right|_Q =
\frac{1}{h_0'(Q)}\frac{\partial h(p,
q)}{\partial p}
\ee
When compared with the Hamiltonian equation for the
original system,
\be
\frac{\partial q}{\partial t} = \frac{\partial h(p,q)}{\partial p},
\ee
we see that $P = h'_0(Q)t$ is proportional to the ordinary time-variable $t$.
This is a usual feature of classical integrable systems, exploited in
Seiberg-Witten theory \cite{GKMMM}:
in the $SU(2)$ case the spectral curve $q(t)$ can be described by
\be
h\left(p\left(\frac{\partial q}{\partial t},q\right),\ q\right) = E.
\label{times}
\ee
where $p$ is expressed through $\partial q/\partial t$ and $q$
from the Hamiltonian equation
$\partial q/\partial t = \partial H/\partial p$.
In other words, the spectral curve is
essentially the solution of the equation of motion of integrable system,
where the time $t$ plays the role of the spectral parameter
and the energy $E$ that of the modulus.

\section{Elliptic Calogero model and its dual}

%\subsection{The dual model}

Here we begin with the elliptic Calogero Hamiltonian
\be
h(p,q) = \frac{p^2}{2} + \frac{g^2}{\sn^2(q|k)},
\ee
and seek a dual Hamiltonian elliptic in the momentum. Thus
$h_0(p)=\frac{p^2}{2}$ and we seek $H(P,Q)=H_0(q)
$ such that $H_0(q) =
\cn(q|k)$. Eqs.(\ref{duality}) become
\be
\frac{Q^2}{2} = \frac{p^2}{2} + \frac{g^2}{\sn^2(q|k)}, \nn \\
\cn(q|k) = H(P,Q), \nn \\
p \cdot \cn'(q|k) = Q\frac{\partial H(P,Q)}{\partial P}.
\label{Calduality}
\ee
Upon substituting
\be
\cn'(q|k) = -\sn(q|k)\dn(q|k) =
%-\sqrt{(1 - \\cn^2(q|k))(k'^2 + k^2\cdot \cn^2(q|k))} = \nn \\ =
-\sqrt{(1 - H^2)(k'^2 + k^2 H^2)},
\label{cnid}
\ee
(this is because $\sn^2 q = 1 - \cn^2 q$,
$\dn^2 q = k'^2 + k^2\cn^2 q$, $k'^2 + k^2 = 1$ and
$\cn q = H$)
we get for (\ref{Calduality}):
%\be
%\sqrt{1 + \frac{2g^2}{Q^2\sn^2(q|k)}}P =
%\int \frac{dH}{\sqrt{(1 - H^2)(k'^2 + k^2 H^2)}}
%\ee
\be
\left(\frac{\partial H}{\partial P}\right)^2 = \frac{p^2}{Q^2}
(1-H^2)(k'^2 + k^2H^2).
\ee
Now from the first eqn.(\ref{Calduality}) $p^2$ can be expressed
through $Q$ and $\sn^2(q|k) = 1 - \cn^2(q|k) = 1 - H^2$ as
\be
\frac{p^2}{Q^2} = 1 - \frac{2g^2}{Q^2(1- H^2)},
\ee
so that
\be
\left(\frac{\partial H}{\partial P}\right)^2
= \left( 1 - \frac{2g^2}{Q^2} - H^2\right)
\left( k'^2 + k^2H^2\right).
\ee
Therefore $H$ is an elliptic function of $P$, namely
\be
H(P,Q) = \cn(q|k) = \alpha(Q) \cdot
\cn\left(P\sqrt{k'^2 + k^2\alpha^2(Q)}\ \bigg| \
\frac{k\alpha(Q)}{\sqrt{k'^2 + k^2\alpha^2(Q)}}\right)
\label{dualCal}
\ee
with
\be
\alpha\sp2(Q) = \alpha\sp2_{rat}(Q) = 1 - \frac{2g^2}{Q^2}.
\ee

In the limit $g^2 = 0$, when the interaction is switched off,
$\alpha(q) = 1$ and $H(P,Q)$ reduces to $H_0(P) =\ \cn(P|k)$,
as assumed in (\ref{Calduality}).

We have therefore obtained a dual formulation of the elliptic Calogero
model (in the simplest $SU(2)$ case). At first glance our dual Hamiltonian
looks somewhat unusual. In particular, the relevant elliptic curve is
``dressed'': it is described by an effective modulus
\be
k_{eff} = \frac{k\alpha(Q)}{\sqrt{k'^2 + k^2\alpha^2(Q)}}
= \frac{k\alpha(Q)}{\sqrt{1 - k^2(1- \alpha^2(Q))}},
\ee
which differs from the ``bare'' one $k$ in a $Q$-dependent way.
In fact $k_{eff}$ is nothing but the
modulus of the ``reduced''
Calogero spectral curve \cite{IM}, see eq.(\ref{keff}) below.

Let us rewrite (\ref{dualCal}) in several equivalent forms.
First, we may solve for $\alpha(Q)$ in terms of $k$ and $k_{eff}$,
\be
\alpha(Q) = \frac{k'k_{eff}}{kk'_{eff}} \ \ \ {\rm and} \ \ \
\beta(Q) \equiv \sqrt{k'^2 + k^2\alpha^2(Q)} =
\frac{k'}{k'_{eff}}.
\ee
Thus (\ref{dualCal}) may be expressed as
\be
H(P,Q) =\cn(q|k) = \frac{k'k_{eff}}{kk'_{eff}} \cn\left(
P\frac{k'}{k'_{eff}}\ \bigg|\ k_{eff}\right),
\ee
from which it follows
\be
\dn(q|k) = \sqrt{k'^2 + k^2\cn^2(q|k)} =
\frac{k'}{k'_{eff}} \dn\left(
P\frac
{k'}{k'_{eff}}\ \bigg|\ k_{eff}\right).
\label{dnnew}
\ee

Interesting expressions arise when we express our results in terms of
of theta-functions. Recall the
standard relations:
\be
\sn (q) = \sqrt{\frac{e_{13}}{\wp(\check q) - e_3}} =
\frac{1}{\sqrt{k}}\frac{\vartheta_1(\hat q)}{\vartheta_4(\hat q)} =
\frac{1}{\sqrt{k}}\frac{\theta_{11}(\hat q|\tau)}
{\theta_{01}(\hat q|\tau)},
\nn \\
\cn (q) = \sqrt{\frac{\wp(\check q) - e_1}{\wp(\check q) - e_3}} =
\sqrt{\frac{k'}{k}}\frac{\vartheta_2(\hat q)}{\vartheta_4(\hat q)} =
\sqrt{\frac{k'}{k}}\frac{\theta_{10}(\hat q|\tau)}
{\theta_{01}(\hat q|\tau)}, \nn \\
\dn (q) = \sqrt{\frac{\wp(\check q) - e_2}{\wp(\check q) - e_3}} =
\sqrt{k'}\frac{\vartheta_3(\hat q)}{\vartheta_4(\hat q)} =
\sqrt{k'}\frac{\theta_{00}(\hat q|\tau)}
{\theta_{01}(\hat q|\tau)}.
\label{ellid}
\ee
Here the Jacobi moduli $k^2$ and $k'^2 = 1 - k^2$
are the cross-ratios of the ramification points of the (hyper-) elliptic
representation of the torus,
\be
y^2 = \prod_{a=1}^3(x - e_a(\tau)), \ \ \
\sum_{a=1}^3 e_a = 0, \ \ \ x = \wp(\check q), \ \ \
y = \frac{1}{2}\wp'(\check q).
\ee
Then
\be
k^2 = \frac{e_{23}}{e_{13}} =
\frac{\vartheta_2^4(0)}{\vartheta_3^4(0)} =
\frac{\theta_{10}^4(0|\tau)}{\theta_{00}^4(0|\tau)}, \ \ \
k'^2 = 1 - k^2 = \frac{e_{12}}{e_{13}} =
\frac{\vartheta_4^4(0)}{\vartheta_3^4(0)} =
\frac{\theta_{01}^4(0|\tau)}{\theta_{00}^4(0|\tau)}
\ee
and
\be
e_{ij} = e_i - e_j, \ \ \
q = 2K\hat q,\ \ \ \check q = 2\omega\hat q, \ \ \
e_{13} = \frac{K^2}{\omega^2}
%\ \ \ \omega = \frac{1}{2}.
\ee
%(The choice $\omega = \frac{1}{2}$ gives $2\pi $ periodicty.)

Similarly,
\be
P = 2K_{eff}\hat P,\ \ \ E_{13} = \frac{K_{eff}^2}{\omega^2_{eff}}
%\ \ \ \omega = \frac{1}{2},
\ee
where $E_1, E_2, E_3$ are the ramification points of the "hyperelliptic"
representation of the ``dressed" torus with modulus $\tau_{eff}$.
This has two equivalent \lq\lq hyperelliptic" representations \cite{IM}:
\be\label{eq}
Y^2=\prod_{a=1}^3(X-E_a)\ \ \ \hbox{and}\ \ \
\hat Y^2=(x-u)\prod_{a=1}^3(x-e_a).
\ee
The equivalence of these representations follows from
the rational map
\be
{u-e_3\over u-e_1}{x-e_1\over x-e_3}={X-E_1\over X-E_3}.
\ee
This allows another interpretation of formula (\ref{dualCal}):
here $x$ is the Weierstrass function related to the elliptic cosine in the
left hand side of (\ref{dualCal}), while $X$ is
that related to the elliptic cosine in the right hand side of (\ref{dualCal})
($Y'$, $Y$ are the first derivatives of the corresponding
Weierstrass functions).
Note that $u$ in (\ref{eq})
is related to the standard Seiberg-Witten modulus by a factor of $2g^2$:
$u=u_{SW}/2g^2$ and to the energy parameter $E$ by $E={u_{SW}-
2g^2e_3\over e_{13}}$. This rescaling factor is responsible for the unusual
coefficient $\sqrt{-2g^2}$ in our definition of the modulus $a$ below (in
(\ref{a})).

The relation between ramification points in the different representations
is easily obtained, since the rational equivalence between the two sets of
points, $e_1, e_2, e_3, u, x, x'$ and
$E_1, E_2, E_3, \infty, X, X'$ implies the following cross-ratio identities for
quadruples:
\be
\frac{E_{ac}}{E_{bc}} = \frac{e_{ac}}{e_{bc}}
\cdot\frac{e_b - u}{e_a -u},\nn\\
\frac{X-E_c}{E_{ac}} =
\frac{x-e_c}{x-u}\cdot\frac{e_a-u}{e_{ac}},\nn\\
\frac{X-X'}{X-E_c}\cdot\frac{E_{ac}}{E_a-X'} =
\frac{x-x'}{x-e_c}\cdot\frac{e_{ac}}{e_a-x'}, \nn\\ i.e.\ \
\frac{E_{ac}dX}{(X-E_a)(X-E_c)} =
\frac{e_{ac}dx}{(x-e_a)(x-e_c)}.
\ee
It then follows that
\be
\frac{dx}{y\sqrt{x-u}} = \frac{dX}{Y}\sqrt{\frac{E_{12}/e_{12}}{e_3-u}} =
\frac{dX}{Y}\sqrt{\frac{E_{23}/e_{23}}{e_1-u}} =
\frac{dX}{Y}\sqrt{\frac{E_{31}/e_{31}}{e_2-u}}.
\ee

Note that with the above
definitions, we find that (\ref{dnnew}) takes the form
\be
\frac{\theta_{00}(\hat q|\tau)}{\theta_{01}(\hat q|\tau)} =
\sqrt{\frac{k'}{k'_{eff}}}
\frac{\theta_{00}(\hat P k'/k'_{eff}|\tau_{eff})}
{\theta_{01}(\hat P k'/k'_{eff}|\tau_{eff})},
\ee
or, more symmetrically,
\be
\frac{\theta_{00}(\hat q|\tau)\theta_{00}(0|\tau)}
{\theta_{01}(\hat q|\tau)\theta_{01}(0|\tau)} =
\frac{\theta_{00}(\hat P k'/k'_{eff}|\tau_{eff})\theta_{00}(0|\tau_{eff})}
{\theta_{01}(\hat P k'/k'_{eff}|\tau_{eff})\theta_
{01}(0|\tau_{eff})}.
\label{thrtheta}
\ee
In fact, after a Landen transformation, we find that
\be
\frac{\theta_{00}(\frac{\hat q}{2}|\frac{\tau}{2})}
{\theta_{01}(\frac{\hat q}{2}|\frac{\tau)}{2}} = \pm
\left(
\frac{\theta_{00}(\frac{\hat P k'}{2k'_{eff}}|\frac{\tau_{eff}}{2})}
{\theta_{01}(\frac{\hat P k'}{2k'_{eff}}|\frac{\tau_{eff}}{2})}
\right)^{\pm 1}.
\label{theta}
\ee

In terms of $P$ and $k_{eff}$ or $\tau_{eff}$ the symplectic
structure $dP\wedge dQ$ looks somewhat more complicated,
but these alternate representations can be useful for other purposes,
including discussion of the algebraic geometry of the spectral curves.

At this stage we have everything to relate the symplectic structure
$dP\wedge dQ$ to the \lq\lq canonical" one, $dp^{Jac}\wedge da$
(see footnote \ref{SWstr}).
First of all, the variation of the flat modulus $a = \oint_A dS$ is
\be\label{a}
da = \g d\left(\oint_A\frac{\sqrt{x-u}}{2y}dx\right) =
-\frac{\g}{4}\left(\oint_A\frac{dx}{y\sqrt{x-u}}\right)du =
-\frac{\g}{4}\sqrt{\frac{E_{13}}{e_{13}}}
\frac{du}{\sqrt{e_2-u}}\oint_A\frac{dX}{Y}.
\ee
Now on the one hand we have
\be\label{keff}
k_{eff}^2 = \frac{E_{23}}{E_{13}} =
\frac{e_{23}}{e_{13}}\cdot\frac{e_1 - u}{e_2 -u} =
k^2\frac{e_1 - u}{e_2 -u}=k^2{E-2g^2\over E-2g^2k^2}
\ee
while on the other hand we have
\be
k_{eff}^2 = \frac{k^2\alpha^2}{\beta^2},
\ee
where
\be
\alpha^2 = 1 - \frac{2g^2}{Q^2}\ \ \ {\rm and} \ \ \
\beta^2 = k'^2 + k^2\alpha^2 = 1 - \frac{2g^2k^2}{Q^2}.
\ee
Thus
\be
\frac{e_1 - u}{e_2 - u} = 1 + \frac{e_{12}}{e_2-u}
= \frac{\alpha^2}{\beta^2} = 1 - \frac{2g^2k'^2}{\beta^2Q^2}
\ee
with $k'^2 = e_{12}/e_{13}$, and so
\be
\frac{e_{13}}{e_2-u} = -\frac{2g^2}{\beta^2Q^2}, \ \ \
du = \frac{e_{13}}{g^2}QdQ.
\ee
Utilizing these gives
\be
da = -\frac{\g}{4}\sqrt{\frac{E_{13}}{e_{13}}}
\frac{du}{\sqrt{e_2-u}}\oint_A\frac{dX}{Y}
= -\g\frac{\sqrt{E_{13}e_{13}}}{4g^2}
\frac{QdQ}{\sqrt{e_2-u}}\oint_A\frac{dX}{Y}
= -\frac{1}{2\beta}dQ\left(\sqrt{E_{13}}\oint_A\frac{dX}{Y}\right).
\ee
Combining these expressions then yields
\be
dP\wedge dQ = 2 d(\beta P)\wedge da
\left(\sqrt{E_{13}}\oint_A\frac{dX}{Y}\right)^{-1}.
\ee
Since the coordinate on the Jacobian differs from the
argument of the Jacobi function by a factor $2\omega_{eff}
\sqrt{E_{13}}$,
\be
P\beta = 2\omega_{eff}\sqrt{E_{13}}\cdot p^{Jac},
\ee
and $\oint_A\frac{dX}{Y}=4\omega_{eff}$,
we finally have
\be
dP\wedge dQ = dp^{Jac}\wedge da.
\ee
Thus our symplectic form is the canonical one.

\subsection{Comment 1. Elliptic solution of Calogero model}

According to the argument of \S\ref{HSW} our Hamiltonian
(\ref{dualCal}) should be simply related to the solution $q(t)$
of the equations of motion of the Calogero Hamiltonian,
which in the case of $SU(2)$ are immediately integrated to give
\be
H^{Cal}\left(\frac{\partial q}{\partial P},q\right) = E
\ee

More explicitly, the equation
\be
\frac{dq}{dt} = \sqrt{E - \frac{2g^2}{\sn^2(q|k)}},
\label{caleq}
\ee
has a solution \cite{Per,GaPer}:
\be
\cn(q|k) = \sqrt{1 - \frac{2g^2}{E}}\cdot
\cn\left(t\sqrt{E - 2g^2k^2}\left|
k\sqrt{\frac{E-2g^2}{E-2g^2k^2}}\right.\right).
\label{calsol}
\label{elcalt}
\ee
This may be derived straightforwardly by
differentiating both sides and applying (\ref{cnid}).
Note that the Calogero equation (\ref{caleq}) and the {\it family}
of Calogero spectral curves are essentially independent of the
value of coupling constant $g^2$: it can be absorbed into
rescaling of moduli (like $E$) and the time-variables (like $t$).

In order to see that (\ref{calsol}) is identical to
(\ref{dualCal}) one needs
to put $E = Q^2$ and make the rescaling $P = h'_0(Q)t = Qt$.
With these substitutions we find that
\be
\sqrt{1 - \frac{2g^2}{E}} = \sqrt{1 - \frac{2g^2}{Q^2}} = \alpha_{rat}(Q),
\ee
and
\be
t\sqrt{E - 2g^2k^2} = P\sqrt{1 - \frac{2g^2k^2}{Q^2}} =
P\sqrt{k'^2 + k^2\alpha_{rat}^2(Q)}, \nn \\
k\sqrt{\frac{E-2g^2}{E-2g^2k^2}} =
k\sqrt{\frac{1-2g^2/Q^2}{1-2g^2k^2/Q^2}} =
\frac{k\alpha_{rat}(Q)}{\sqrt{k'^2 + k^2\alpha_{rat}^2(Q)}}.
\ee
We then see that (\ref{calsol}) is identical to (\ref{dualCal}).

We remark that the relevant symplectic structure here is\footnote{
In the following sections we prove that not only the
elliptic-rational (the dual of the elliptic Calogero model) but also
the elliptic-trigonometric (the dual of the elliptic Ruijsenaars model)
and the elliptic-elliptic (our new double-elliptic) Hamiltonians
have the same form (\ref{elcalt}), but the latter with the identifications
$E = \sinh^2Q$ and $E = \sn^2(Q|\tilde k)$.
Thus they are also related to {\it Calogero} equation
(\ref{caleq}). However, the relevant symplectic structures
-- which are always given by $dP\wedge dQ = h'_0(Q)dt\wedge
dQ = dh_0(Q)\wedge dt$ -- are no
longer equivalent to $dE\wedge dt$ (since $E \neq h_0(Q)$, i.e. $E$ is no
longer associated with the proper Hamiltonian). }
\be
dE\wedge dt =
2QdQ\wedge dt = -2dP\wedge dQ.
\ee

\subsection{Comment 2. Projection method}

Another important remark about the Calogero model is that
%for particular value of the coupling constant $g^2$,
its elliptic solution -- and thus
the $SU(2)$ dual Hamiltonian (\ref{dualCal}) -- can be obtained
by the projection method: the spectral curve $q(P)$ is embedded into
its Jacobian (an abelian variety, i.e. a torus of complex dimension
$N = 2$) by a simple algebraic equation
\be
\hat\Theta(\hat q,\hat P)
\equiv \Theta(\hat q+\hat P,\hat q-\hat P)
= 0
\label{projeq}
\ee
where $\hat q$ and $\hat P$ are just the two coordinates on the
Jacobian (hats appear because of the difference in normalization
of arguments of theta and Jacobi elliptic functions, see (\ref{ellid})).
Now the Calogero spectral curve -- and consequently
the relevant genus-two theta-function -- has a very particular
period matrix: the sum of all the elements in every row is the same
(independent of the number of a row),
\be
\sum_{j=1}^N T_{ij} = const
\label{Tpecul}
\ee
In the case of $N=2$ this means
that the period matrix has
$T_{11} = T_{22}$.
The corresponding theta-functions are then easily represented in terms
of genus-one theta-functions. For example:
\be
\hat\Theta(\xi_+,\xi_-) \equiv
\Theta\left[\begin{array}{cc} 1 & 1 \\ 1 & 1 \end{array}\right]\left(
\begin{array}{cc} r & s \\ s & r\end{array}\right)(\xi_1,\xi_2)
= \nn \\ =
\sum_{m,n \in Z} \exp\ i\pi\left(
(m+\frac{1}{2})^2r + 2(m+\frac{1}{2})(n+\frac{1}{2})s +
(n+\frac{1}{2})^2r + 2(m+\frac{1}{2})(\xi_1 + \frac{1}{2}) +
2(n+\frac{1}{2})(\xi_2 + \frac{1}{2})
\right)
\nn \\ =
\sum_{m,n \in Z} \exp\ i\pi\left(
(m+n+1)^2\frac{\tau}{2} + (m-n)^2\frac{\tau_{eff}}{2} +
2(m+n+1)(\xi_+ + \frac{1}{2}) + 2(m-n)\xi_-
\right)
\nn \\ =
\left(\sum_{k \in Z,\ l \in \frac{1}{2} + Z} -
\sum_{k \in \frac{1}{2} + Z,\ l \in Z}\right)
\exp\ i\pi\left(
2k^2{\tau} + 2l^2{\tau_{eff}} +
4k\xi_+ + 4l\xi_-
\right)
\nn \\ =
\theta_{00}(2\xi_+|2\tau)\theta_{10}(2\xi_-|2\tau_{eff}) -
\theta_{10}(2\xi_+|2\tau)\theta_{00}(2\xi_-|2\tau_{eff})
\nn \\ =
\frac{1}{2}\left(
\theta_{01}(\xi_+|\frac{\tau}{2})
\theta_{00}(\xi_-|\frac{\tau_{eff}}{2}) -
\theta_{00}(\xi_+|\frac{\tau}{2})
\theta_{01}(\xi_-|\frac{\tau_{eff}}{2})\right)
\label{decomp}
\ee
where $\tau = r+s$ and $\tau_{eff} = r-s$ and
$\xi_{\pm} = \frac{1}{2}(\xi_1 \pm \xi_2)$.

We see that (after the appropriate $\tau$- and $\tau_{eff}$-dependent
rescaling of $\hat P$) the equation $\hat\Theta(\hat q,\hat P) = 0$
has (\ref{theta}) and thus (\ref{dualCal}) as a solution.
The different choices of sign in (\ref{theta}) correspond to
different choices of theta-characteristics in
(\ref{projeq}) and (\ref{decomp}), and these are related
by modular transformations.
This result follows from the work of \cite{Krisol}; see also
\cite{Krirev,GaPer}\footnote{
In \cite{GaPer} the period matrix is taken to be \cite{B}:
$\left(
\begin{array}{cc} \tau/2 & 1/2 \\ 1/2 & -\tau_{eff}/2
\end{array}\right)$,
which is modular equivalent, but
different from our choice in (\ref{decomp}),
$\left(
\begin{array}{cc} r & s \\ s & r\end{array}\right)$.
The obvious choice of cycles and holomorphic differentials
in the case of Calogero curve, which is the double-covering
of the bare torus $(0,\tau)$, is:
$$ \oint_{A_1} d\omega_\pm = \pm \oint_{A_2} d\omega_\pm = 1$$
and
$$ \oint_{B_1} d\omega_+ = \oint_{B_2} d\omega_+ = \tau, \ \ \
\oint_{B_1} d\omega_- = -\oint_{B_2} d\omega_- = \tau_{eff} $$
Canonical differentials for such cycles are
$d\omega_1 = \frac{d\omega_+ + d\omega_-}{2}$ and
$d\omega_2 = \frac{d\omega_+ - d\omega_-}{2}$
and the period matrix is our $\left(
\begin{array}{cc} r & s \\ s & r\end{array}\right)$.
For another choice of cycles \cite{B}
$\hat A_1 = A_1 + A_2,\ \hat A_2 = B_1-B_2,\ \hat B_1 = B_1,\
\hat B_2 = A_2$ the period matrix
(defined from the relations
$\oint_{B_i}d\omega = T_{ij}\oint_{A_j}d\omega$ for
arbitrary holomorphic $d\omega$) is $\left(
\begin{array}{cc} \tau/2 & 1/2 \\ 1/2 & -1/2\tau_{eff}
\end{array}\right)$.
}.

The projection method provides the most direct generalisation
from $SU(2)$ to $SU(N)$, i.e. to the $N$-particle
systems, described (in the center of mass) by $g=N-1$
independent coordinates and momenta.
Because of (\ref{Tpecul}) the genus-$N$ theta function on the
Jacobian of the Calogero spectral curve always decomposes into
bilinear combinations of genus-one and genus-$g$ theta
functions:
\be
\hat\Theta^{(N)}(\hat q,\hat{\vec P}) = \frac{1}{N}
\sum_{i=1}^N \theta_{*}\left(\hat q + \frac{i}{N}\left|
\frac{\tau}{N}\right.\right)
\Theta^{(g)}_{e_i}\left(\hat{\vec P}\left|
\frac{T_{eff}^{(g)}}{N}\right.\right).
\label{gendecomp}
\ee
The equation
\be
\hat\Theta^{(N)}(\hat q,\hat{\vec P}) = 0
\label{thetagen}
\ee
then defines $N$ branches of the solution $q_i(\vec P)$,
which generalises (\ref{dualCal}).
As usual in Seiberg-Witten theory (see eq.(\ref{times}))
the dual momenta $\vec P$ may be associated with the first $g$
time-variables while the dual coordinates are the moduli, parameterising
the period matrix $T^{(g)}$ which characterises the covering of
the bare elliptic curve.

Alternatively $\hat\Theta^{(N)}(\hat q,\hat{\vec P})$ may be considered
as a generating function for the dual Hamiltonians, with the
original coordinates $q_i$ playing the role of the spectral
parameter (which carries an index $i$, labeling the sheet
of the $N$-sheet covering).

The Hamiltonians themselves are made from the genus-$g$
theta functions
$\Theta^{(g)}_{e_i}\left(\hat{\vec P}\left|
\frac{T^{(g)}}{N}\right.\right)$
with $N$ different theta-characteristics $e_i$ (as (\ref{dualCal})
in the case of $SU(2)$ is made from two genus-one theta-functions
with half-integer characteristics -- which form an elliptic cosine).
The Seiberg-Witten symplectic structure defines the Poisson
bracket between $\vec P$ and $T_{eff}^{(g)}$ such that the Hamiltonians
are Poisson-commuting.
Commutativity is implied by the claim \cite{Krisol,Krirev,GaPer} that
for any $N$ (and, at least, at a special value of the coupling constant
$g^2$) in addition to the explicit decomposition (\ref{gendecomp})
there is also an implicit (at todays level of knowledge)
one into elliptic (genus-one) theta (sigma)-functions:
\be
\hat\Theta^{(N)}(\hat q,\hat{\vec P}) \sim \prod_{i=1}^N
\theta_*\left(\hat q - \hat q_i(\vec P)\ \bigg|
\ \tau\right).
\label{Kridecomp}
\ee
Here $q_i(\vec P|\tau, T_{eff}^{(g)})$ are coordinates of the $SU(N)$ Calogero
equations and so, for a given set of $g$ times $\vec P$, these do
Poisson-commute.
Note that the non-trivial coefficient of proportionality in
(\ref{Kridecomp}) means the $\tau$-functions
$\hat\Theta^{(N)}(\hat q,\hat{\vec P})$ need not commute
at different values of the spectral parameter $q$. In particular, the
individual coefficients
$\Theta^{(g)}_{e_i}\left(\hat{\vec P}\left|
\frac{T^{(g)}}{N}\right.\right)$
also need not commute\footnote{
For example, in the case of $N=2$,
$$
\left\{\theta_{00}(\hat P |\frac{T}{2}),\
\theta_{01}(\hat P |\frac{T}{2})\right\} =
\frac{i}{4\pi} \left\{P,\frac{T}{2}\right\}\cdot
(\theta_{00}'\theta_{01}'' - \theta_{00}''\theta_{01}')
(P|\frac{T}{2})
\neq 0
$$
}
and only their ratios will. These ratios form the
Poisson-commuting Hamiltonians.
%It can be convenient to look for explicit form of these Hamiltonians
%in explicit parametrizations of the spectral curves, starting,
%for example, from the hyperelliptic representation,
%$y^2 = (x-e_1)(x-e_2)(x-e_3)$ of the bare curve.

In order to get a double-elliptic system one needs to change the
parameterisation of $T^{(g)}$ from rational to elliptic, or,
equivalently, to adequately deform the Seiberg-Witten
symplectic structure. In section 5 below we present such a
deformation for the $SU(2)$ case.

\section{Elliptic Ruijsenaars System}

All of the above formulae are straightforwardly generalised
from the Calogero (rational-elliptic) system to the Ruijsenaars
(trigonometric-elliptic) system. The only difference ensuing is that
the $q$-dependence of the dual (elliptic-trigonometric) Hamiltonian
is now trigonometric rather than rational:
\be
\alpha\sp2(q) = \alpha\sp2_{trig}(q) = 1 - \frac{2g^2}{\sinh^2 q}
\ee
For details on the geometry of the Ruijsenaars spectral
curves see \cite{BMMM2}.
Rather than giving further details we will proceed directly to
a consideration of the double-elliptic model.

\section{The double-elliptic system}

\subsection{Solution of duality equations}

In order to get a double-elliptic system one needs to
exchange the rational $Q$-dependence
in (\ref{Calduality}) for elliptic an one, and so we substitute
$\alpha\sp2_{rat}(Q)$ by the obvious elliptic
analogue $\alpha\sp2_{ell}(Q) = 1 - \frac{2g^2}{\sn^2(Q|\tilde k)}$.
Moreover, now the elliptic curves for $q$ and $Q$ need not in general
be the same, i.e. $\tilde k \neq k$.

Instead of (\ref{Calduality}) the duality equations now become
\be
\cn(q|k) = H(P,Q|k,\tilde k), \nn \\
\cn(Q|\tilde k) = H(p,q|\tilde k,k), \nn \\
\cn'(Q|\tilde k)\frac{\partial H(P,Q|k,\tilde k)}{\partial P} =
\cn'(q|k)\frac{\partial H(p,q|\tilde k,k)}{\partial p},
\label{dellduality}
\ee
and the natural ansatz for the Hamiltonian (suggested
by (\ref{dualCal})) is
\be
H(p,q|\tilde k,k) = \alpha(q|\tilde k,k)\cdot
\cn\left(p\;\beta(q|\tilde k,k)\ |\ \gamma(q|\tilde k,k)\right)
= \alpha\cn(p\beta|\gamma), \nn \\
H(P,Q|k,\tilde k) = \alpha(Q|k,\tilde k)\cdot
\cn\left(P\;\beta(Q|k,\tilde k)\ |\ \gamma(Q|k,\tilde k)\right)
= \tilde\alpha \cn(P\tilde\beta|\tilde\gamma).
\label{dellans}
\ee
For ease of expression we will suppress the dependence of
$\alpha,\beta,\gamma$ on $k$ and $\tilde k$ in what follows
using $\alpha(q)$ for $\alpha(q|\tilde k,k)$
and $\tilde\alpha(Q)$ for $\alpha(Q|k,\tilde k)$ etc.

Substituting these ansatz into (\ref{dellduality}) and making use
of (\ref{cnid}), the square of the final eqn.(\ref{dellduality}) becomes
\be
\left(1 - \cn^2(Q|\tilde k)\right)
\left(\tilde k'^2 + \tilde k^2\cn^2(Q|\tilde k)\right)
\tilde\alpha^2(Q)\tilde\beta^2(Q)
\left(1 - \cn^2(P\tilde\beta|\tilde\gamma)\right)
\left(\tilde \gamma'^2 +
\tilde\gamma^2\cn^2(P\tilde\beta|\tilde\gamma)\right)
\nn \\ =
\left(1 - \cn^2(q|k)\right)
\left(k'^2 + k^2\cn^2(q|k)\right)
\alpha^2(q)\beta^2(q)
\left(1 - \cn^2(p\beta|\gamma)\right)
\left(\gamma'^2 +
\gamma^2\cn^2(p\beta|\gamma)\right)
\ee
The first two eqs.(\ref{dellduality}) together with (\ref{dellans})
allow this to be simplified yielding
\be
\tilde\beta^2(Q)\left(1 - \cn^2(Q|\tilde k)\right)
\left(\tilde k'^2 + \tilde k^2\cn^2(Q|\tilde k)\right)
\left(\tilde\alpha^2(Q) - \cn^2(q|k)\right)
\left(\tilde \gamma'^2 +
\frac{\tilde\gamma^2}{\tilde\alpha^2}\cn^2(q|k)\right)
= \nn \\ =
\beta^2(q)\left(1 - \cn^2(q|k)\right)
\left(k'^2 + k^2\cn^2(q|k)\right)
\left(\alpha^2(q) - \cn^2(Q|\tilde k)\right)
\left(\gamma'^2 +
\frac{\gamma^2}{\alpha^2}\cn^2(Q|\tilde k)\right)
\label{prom1}
\ee

Now there is cancellation between the third and fifth terms of the
left and right hand sides provided
\be
\frac{\tilde k'^2}{\tilde k^2} =
\frac{\alpha^2\gamma'^2}{\gamma^2}(q),
\nn \\
\frac{k'^2}{k^2} =
\frac{\tilde\alpha^2\tilde\gamma'^2}{\tilde\gamma^2}(Q).
\ee
Then, since $\gamma'^2 \equiv 1 - \gamma^2$, these may be reexpressed
as
\be
\gamma^2(q) = \frac{\tilde k^2\alpha^2(q)}
{\tilde k'^2 + \tilde k^2\alpha^2(q)}, \nn \\
\tilde\gamma^2(q) = \frac{k^2\tilde\alpha^2(Q)}
{k'^2 + k^2\tilde\alpha^2(Q)}.
\ee
With these identifications we now obtain from (\ref{prom1}) that
\be
\frac{\tilde\beta^2(Q)}{k'^2 + k^2\tilde\alpha^2(Q)}
\left( 1 - \cn^2(Q|\tilde k) \right)
\left(\tilde\alpha^2(Q) - \cn^2(q|k)\right) = \nn \\ =
\frac{\beta^2(q)}{
\tilde k'^2 + \tilde k^2\alpha^2(q)}
\left(1 - \cn^2(q|k)\right)
\left((\alpha^2(q) - \cn^2(Q|\tilde k)\right).
\ee
This relation should hold for all values of the two independent
variables $q$ and $Q$. These variables can be separated
provided the terms $\cn^2(q|k)\cdot \cn^2(Q|\tilde k)$ on both
sides cancel each other. This implies that
\be
\beta^2(q) = \tilde k'^2 + \tilde k^2\alpha^2(q), \nn \\
\tilde\beta^2(Q) = k'^2 + k^2\tilde\alpha^2(Q),
\ee
and
\be
\left( 1 - \cn^2(Q|\tilde k) \right)
\left(\tilde\alpha^2(Q) - \cn^2(q|k)\right) =
\left(1 - \cn^2(q|k)\right)
\left((\alpha^2(q) - \cn^2(Q|\tilde k)\right)
\ee
i.e.
\be
\alpha^2(q) \sn^2(q|k) + \cn^2(q|k) =
\tilde\alpha^2(Q|\tilde k) \sn^2(Q|\tilde k) +
\cn^2(Q|\tilde k) =
1 - 2g^2 = const.
\ee
Here we have represented the $q$ and $Q$-independent constant as
$1 - 2g^2$ to introduce the coupling constant $g^2$ in the conventional manner.
Thus we arrive at
\be
\alpha^2(q|\tilde k,k) = \alpha^2(q|k) =
1 - \frac{2g^2}{\sn^2(q|k)}, \nn \\
\beta^2(q|\tilde k,k) = \tilde k'^2 + \tilde k^2\alpha^2(q|k), \nn \\
\gamma^2(q|\tilde k,k) = \frac{\tilde k^2\alpha^2(q|k)}
{\tilde k'^2 + \tilde k^2\alpha^2(q|k)},
\ee
and finally the double-elliptic duality becomes
\be
H(P,Q|k,\tilde k)= \cn(q|k) =
\alpha(Q|\tilde k) \cn\left(
P\sqrt{k'^2 + k^2\alpha^2(Q|\tilde k)}\ \bigg|\
\frac{k\alpha(Q|\tilde k)}{\sqrt{k'^2 + k^2\alpha^2(Q|\tilde k)}}
\right),
\ee
\be
H(p,q|\tilde k,k) = \cn(Q|\tilde k) =
\alpha(q|k) \cn\left(p\sqrt{\tilde
k'^2 + \tilde k^2\alpha^2(q|k)}\ \bigg|\
\frac{\tilde k\alpha(q|k)}
{\sqrt{\tilde k'^2 + \tilde k^2\alpha^2(q|k)}}
\right).
\ee

These double-elliptic Hamiltonians are our main new result. We shall
now consider various limiting cases arising from these, and discuss
various other choices that can be made as an ansatz.

\subsection{Limiting cases}

We now show that the double-elliptic Hamiltonian
\begin{equation}
\begin{array}{rl}
H^{dell}(p,q|\tilde k,k) &\equiv
\alpha(q|k) \cdot \cn\left(
p\sqrt{\tilde k'^2 + \tilde k^2\alpha^2(q|k)}\left|
\frac{\tilde k\alpha(q|k)}
{\sqrt{\tilde k'^2 + \tilde k^2\alpha^2(q|k)}}
\right.\right),\\
\alpha^2(q|k) &= 1 - \frac{2g^2}{\sn^2(q|k)},
\end{array}
\label{dellHam}
\end{equation}
contains the entire Ruijsenaars-Calogero and Toda family
as its limiting cases, as desired. (Of course we have restricted ourselves
to the $SU(2)$ members of this family in this paper.)

In order to convert the elliptic dependence of the momentum $p$
into the trigonometric one, the corresponding ``bare''
modulus $\tilde k$ should vanish: $\tilde k \rightarrow 0, \
\tilde k'^2 = 1 - \tilde k^2 \rightarrow 1$ (while $k$
can be kept finite). Then, since
$\cn(x|\tilde k = 0) = \cosh x$,
\be
H^{dell}(p,q) \longrightarrow \alpha(q)\cosh p =
H^{Ru}(p,q)
\ee
with the same
\be
\alpha^2(q|k) = 1 - \frac{2g^2}{\sn^2(q|k)}.
\ee
Thus we obtain the $SU(2)$ elliptic Ruijsenaars Hamiltonian.\footnote{
Indeed,
$$
F^2(q) = c^2(\check\epsilon|k)
\left(\wp(\check\epsilon) - \wp(\check q)\right)
= c^2(\check\epsilon|k)\left(\frac{e_{13}}
{\sn^2(\sqrt{e_{13}}\check \epsilon|k)} -
\frac{e_{13}}{\sn^2(\sqrt{e_{13}}\check q|k)}\right)
=
\frac{c^2(\check\epsilon|k) e_{13}}{\sn^2(q|k)}
\left(1 - \frac{\sn^2(\epsilon|k)}{\sn^2(q|k)}\right)
$$
where $q = 2\omega\hat q\sqrt{e_{13}}$
and $2g^2 = \sn^2(\epsilon|k)$.
}
The trigonometric and rational Ruijsenaars
as well as all of the Calogero and
Toda systems are obtained through further limiting procedures
in the standard way.

The other limit $k \rightarrow 0$ (with $\tilde k$ finite) gives
$\alpha(q|k) \rightarrow \alpha_{trig}(q) = 1 - \frac{2g^2}{\cosh q}$
and
\begin{equation}
\begin{array}{rl}
H^{dell}(p,q) \longrightarrow &
\alpha_{trig}(q)\cdot
\cn\left(
p\sqrt{\tilde k'^2 + \tilde k^2\alpha_{trig}^2(q)}\left|
\frac{\tilde k\alpha_{trig}(q)}
{\sqrt{\tilde k'^2 + \tilde k^2\alpha_{trig}^2(q)}}
\right.\right) = \tilde H^{Ru}(p,q).\\
\end{array}
\end{equation}
This is the elliptic-trigonometric model, dual to the
conventional elliptic Ruijsenaars (i.e. the trigonometric-elliptic)
system.
In the further limit of small $q$
this degenerates into the elliptic-rational
model with $\alpha_{trig}(q) \rightarrow \alpha_{rat}(q) =
1 - \frac{2g^2}{q^2}$, which is dual to the conventional elliptic
Calogero (i.e. the rational-elliptic) system, analysed in some detail
in section three above.

\subsection{Other double-elliptic ansatze}

Our approach has been based on choosing appropriate functions
$f(q)$ and $F(Q)$ and implementing duality. Other choices of
functions associated with alternative free Hamiltonians may be possible.
Instead of the duality relations (\ref{dellduality}) one could consider
those based on $h_0(p) = \sn(p|\tilde k)$ instead of $\cn(p|\tilde k)$.
These give
\be
\sn(q|k) = H_s(P,Q|k,\tilde k), \nn \\
\sn(Q|\tilde k) = H_s(p,q|\tilde k,k), \nn \\
\sn'(Q|\tilde k)\frac{\partial H_s(P,Q|k,\tilde k)}{\partial P} =
\sn'(q|k)\frac{\partial H_s(p,q|\tilde k,k)}{\partial p},
\label{dellduality1}
\ee
and now the natural ansatz is
\be
H_s(p,q|\tilde k,k) = \alpha_s(q|\tilde k,k)\cdot
\sn\left(p\beta_s(q|\tilde k,k)\ |\ \gamma_s(q|\tilde k,k)\right).
\label{dellansin}
\ee
All of our calculations above may be repeated with little difference,
the only significant one being that instead of (\ref{cnid}) one now uses
\be
\sn'(q|k) = \sqrt{(1 - \sn^2(q|k))(1- k^2\sn^2(q|k))}.
\ee
With this choice one gets somewhat simpler expressions for $\beta_s$
and $\gamma_s$:
\be
\beta_s = 1, \nn \\
\gamma_s(q|\tilde k,k) = \tilde k\alpha_s(q|k), \\
\alpha_s(q|k) = 1 - \frac{2g^2}{\cn^2(q|k)}
\ee
and the final Hamiltonian is now
\be
H_s(p,q|\tilde k,k) = \alpha_s(q|k) \cdot
\sn(p|\tilde k\alpha_s(q|k)).
\label{dellHamsin}
\ee
Although this Hamiltonian is somewhat simpler than our earlier choice,
the limits involved in obtaining the Ruijsenaars-Calogero-Toda
reductions are somewhat more involved, and that is why we
chose to present the Hamiltonian (\ref{dellHam}) first.

One might further try other elliptic functions for $h_0(p)$.
Every solution we have obtained by making a different ansatz
has been related to our solution (\ref{dellHam}) via modular transformations
of the four moduli $\tilde k$, $k$, $\tilde k_{eff} =
\tilde\gamma$ and $k_{eff} = \gamma$.

\section{Conclusion}

In this paper
we suggest that the 2-particle ($SU(2)$) Hamiltonian of the
double-elliptic system is given by:
\be
H(p,q|\tilde k,k) = \alpha(q|k) \cn\left(
p\sqrt{\tilde k'^2 + \tilde k^2\alpha^2(q|k)}\ \bigg|\
\frac{\tilde k\alpha(q|k)}
{\sqrt{\tilde k'^2 + \tilde k^2\alpha^2(q|k)}}
\right),
\ee
\be
\alpha^2(q|k) = 1 - \frac{2g^2}{\sn^2(q|k)}
\ee
As particular limits this model provides the elliptic Ruijsenaars system
($\tilde k\rightarrow 0$) and its dual ($k\rightarrow 0$).

A non-trivial feature of our double-elliptic model is the ``dressing''
of the bare elliptic moduli $\tilde k$ and $k$ which characterise
the momentum and coordinate tori respectively.
In general the geometry of the double-elliptic system involves
two elliptic curves and two ``dressed'' Jacobians
($N-1$-dimensional algebraic varieties)
with the period matrices $\tilde\tau$, $\tau$,
$\tilde T^{(g)}_{eff}$ and $T^{(g)}_{eff}$.
The spectral parameters can be associated with the
center-of-mass momentum and coordinate or, equivalently,
with theta-characteristics, as implicitly explained in
\cite{BMMM2}.
The multi-particle generalisation, introduction of spectral
parameters, and the relationship between these models and
Yang-Mills in $6d$,
to double-loop algebras and conformal models,
and to $\tau$-functions and their fermionic representations
will be discussed elsewhere.

\section{Acknowledgements}

We are indebted to V.Fock,
A.Gerasimov, A.Gorsky, S.Kharchev, I.Krichever,
M.Olshanetsky and A.Zabrodin
for numerous discussions on the subject of this paper.

Our work is partly supported by
RFBR grants 98-01-0344 (A.Mar.), 98-01-0328 (A.Mir.),
98-02-16575 (A.Mor.), INTAS grants 96-482 (A.Mar.), 97-0103 (A.Mir.)
and Russian President's grant 96-15-96939 (A.Mor.).
A.Mir. and H.W.B. also acknowledge the Royal Society for support under a
joint project.

\end{document}